\documentstyle[12pt]{article}
\font\blackboard=msbm10 at 12pt
\font\blackboards=msbm7
\font\blackboardss=msbm5
\newfam\black
\textfont\black=\blackboard
\scriptfont\black=\blackboards
\scriptscriptfont\black=\blackboardss
\def\bb#1{{\fam\black\relax#1}}
\newlength{\dinwidth}
\newlength{\dinmargin}
\begin{document}
\newcommand{\z}{{\bb Z}}
\newcommand{\r}{{\bb R}}
\newcommand{\bc}{{\bb C}}
\newcommand{\be}{\begin{equation}}
\newcommand{\ee}{\end{equation}}
\newcommand{\ber}{\begin{eqnarray}}
\newcommand{\eer}{\end{eqnarray}}
\newcommand{\lp}{\left(}
\newcommand{\rp}{\right)}
\newcommand{\lk}{\left\{}
\newcommand{\rk}{\right\}}
\newcommand{\lc}{\left[}
\newcommand{\rc}{\right]}
\newcommand{\sT}{{\scriptscriptstyle T}}
\newcommand{\2}{\,\,2}
\def\a{\alpha}
\def\b{\beta}
\def\g{\gamma}
\newcommand{\se}{\section}
\newcommand{\tra}{\vec{p}_{\sT}}
\newcommand{\Z}{Z\left(\beta\right)}
\newcommand{\half}{\frac{1}{2}}
\thispagestyle{empty}

\begin{flushright}
\begin{tabular}{l}
FFUOV-98/11\\
{\tt hep-th/9811170}\\
\today
\end{tabular}
\end{flushright}

\vspace*{2cm}

{\vbox{\centerline{{\Large{\bf MATRIX STRINGS, COMPACTIFICATION SCALES
}}}}}
\vspace*{1cm}
{\vbox{\centerline{{\Large{\bf AND
HAGEDORN TRANSITION
}}}}}   
\vskip30pt

\centerline{Marco Laucelli Meana and Jes\'{u}s Puente Pe\~{n}alba
\footnote{laucelli@string1.ciencias.uniovi.es \\
jesus@string1.ciencias.uniovi.es}}

\vskip6pt
\centerline{{\it Dpto. de F\'{\i}sica, Universidad de Oviedo}}
\centerline{{\it Avda. Calvo Sotelo 18}}
\centerline{{\it E-33007 Oviedo, Asturias, Spain}}

\vskip .5in

\begin{center}
{\bf Abstract}
\end{center}

In this work we use the Matrix Model of Strings in order to extract some
non-perturbative information on how the Hagedorn critical temperature
arises from eleven-dimensional physics. We study the thermal behavior of
M and Matrix theories on the compactification backgrounds that correspond to 
string models. We obtain some information that allows us to state that the
Hagedorn temperature is not unique for all Matrix String models and we are
also able to sketch how the $S$-duality transformation works in this
framework.

\newpage

\section{Introduction}

The Matrix Model of M-Theory  \cite{BFSS,SUS,SEI} has been shown to include the second
quantized string spectrum
\cite{BSE,motl,DVV,rey1,rey2,hethorava,BM,lowe,silv,cv}. In particular it is possible 
to see that  the light-cone description of Type IIA and Heterotic $E_8\times E_8$ string
theories can be obtained through the compactification of the Matrix Model on a
compact manifold, which corresponds to a torus, 
${\bf T}^2=S^1\times S^1$, and the orbifold  $S^1\times S^1/\z_2$ respectively. We can
parameterize these backgrounds with the radii $R$ and ${R_9}$
of the compact dimensions. One of them is taken to be 
the eleven-dimensional quantum length, related to strong coupling dynamics of string 
theory as in \cite{various,horava}. Type I string theories are also contained 
as an orientifold of the type II ones \cite{orbi1,orbi2}.

The idea that allows to obtain string theories from the Matrix Model consist
in taking a membrane configuration of $N$ D-particles of the Matrix
Model in a concrete compactification background. Different limits on the
radii that define the classical moduli lead to different string theories.
It is necessary to make the usually called 'FLIP'. It simply consists in 
the  interchange of the direction we consider the quantum one in such a way
that the light-cone description of string models is obtained.  

This new framework allows us to attempt to understand an old problem of
perturbative string theory. When we study the thermal behavior of a gas of
fundamental strings we arrive at the so-called Hagedorn problem (see
\cite{ALV, OSO1,atick,us1, us2} and references therein). It consists of the
presence of a critical temperature in which the canonical description of the
gas breaks down. The origin of this singularity is  the exponential growth of 
the number of states  at each  mass level of the strings. After an ample
discussion along the last ten years one can conclude that the Hagedorn 
temperature is a maximum one and in the case of open universes it is reached after 
a phase transition \cite{us1,us2}. What is still lacking is a deeper
understanding of what the nature of the phase transition is. We know that, at
least from a perturbative point of view, it seems  that at temperatures near 
the critical one a redistribution of the degrees of freedom of the theory 
is needed. 

One could also wonder if non-perturbative effects would change the
picture. Some works in this direction have been carried out concluding that
these effects do not affect the critical behavior \cite{green,barbovaz}. We
think that it is due to the fact that non-perturbative effects in string
theory are relevant at scales smaller than the string one \cite{bachas,DKPS}, 
and so they could not affect a feature  that naturally arises  at $\alpha'$ 
scales.
 
From the moment we know that string theories are corners of the moduli space
of an eleven-dimensional fundamental theory,  it seems mandatory to try 
to understand what the Hagedorn critical temperature means from an 
M-theoretical perspective. 

In the limit in which the spatial compactification radius, $R_9$, goes to zero
the  membranes of Matrix theory become strings, and then the Hagedorn 
temperature does appear. Some physical properties of this phenomenon have been analyzed in \cite{Sathi}. An interesting feature
of the Hagedorn transition in String Theory  is that it takes different
values depending on the model we analyze. More concretely, the gas of Heterotic 
and Type II strings do not present 
the phase transition at the same temperature.  The idea we will develop
in next sections is that we can extract some information from the Matrix
model compactifications that will allow us to draft some properties of the
thermal behavior of Matrix String Theories, and clarify some aspects of the 
Hagedorn phenomenon from a more fundamental perspective.    

Let us sketch the idea. We will argue  that the  critical thermal properties of 
string theories
 are functions of a ten-dimensional parameter ${\lambda}$ which could be expressed 
in terms of eleven-dimensional variables as $\lambda=\beta R_{9}^{1/2}$.
If we assume the Hagedorn temperature to be unique from the Matrix Model point 
of view, and $R_9$ to be the order parameter of the transition \cite{Sathi}, 
it would be possible to establish a relation between 
the different Hagedorn temperatures of string theories. If we call the transition points 
of the perturbative string gas $\lambda_H$, then the relationship between different Hagedorn
 temperatures
could be obtained by 
\be
\frac{\lambda^{IIA}_{H}}{\lambda^{Het}_{H}}=\lp\frac{ \beta_{H}
\sqrt{R_{9}^{IIA}}}{ \beta_{H}
\sqrt{R_{9}^{Het}}}\rp \stackrel{?}{=}\lp \frac{
R_{9}^{IIA}}{R_{9}^{Het}} \rp^{1/2}.
\label{relation} 
\ee
We will use this approach to elucidate whether the Hagedorn temperature
 is unique in the M-theory framework.

Some words about the open string case are in need.  We know type I string theories have the same 
Hagedorn temperatures as type II ones \cite{ALV,OSO1}. This 
property is easily understood remembering that type I string theories may be 
obtained as a $\z_2$ orientifold of closed strings. In perturbative string
theory the $\z_2$ projection does not change the critical point \cite{ALVOSO}. 
We will argue that from an eleven  dimensional point of view this is to be 
expected, because of the fact that the compact manifold that defines the string 
scale, in this case a circle, is the same. That is the properties of the
quantum dimension determines the value of the critical temperature.

Another approach that could give some information on how  the
compactification background enters into the thermal properties of string
theories is based on the study of the thermalized maximal SUGRA in eleven 
dimensions in those backgrounds. In next section we will show that the
thermal properties of the effective supergravities, corresponding to low
energy limits of string theories, are functions  of the
eleven-dimensional temperature  and the compactification scale. 


\section{M-Theory   origin of string theories}

Historically  M-theory appeared as the theory that would fill the lack of an 
underlying framework in which to understand the  map of string dualities. 
In \cite{various,horava} the eleven-dimensional origin of effective string 
theories was described. As argued in the introduction  we think that we can
obtain some information relating the thermal behavior of string theories and
the eleven-dimensional compactification scale, by studying the M-theoretical 
origin of effective supergravities. This is the scope of this section.

\subsection{M-theory on a torus} 

In \cite{various} was shown that the strong coupling limit of type IIA
String Theory  at low energies is eleven-dimensional SUGRA on $R^{10}\times
S^1$, where the radius of the eleven-dimensional circle is related to the
string coupling constant by $R=l_{11} g^{2/3}_s$.

This relationship between M and IIA theories can be generalized to an
arbitrary  manifold in such a way that we can make the following  statement:
Type IIA String Theory  on a manifold ${\cal M}$ at strong coupling and at
low energies is described by M-theory on ${\cal M}\times S^1$. 

We will use this property to study some interesting background manifolds. An 
example is type IIB case. We know that by $T$-duality we can obtain Type IIB 
string as  the dual of IIA one. More explicitly, taking 
${\cal M}=R^{9}\times S^1$ and acting with the duality on this extra $S^1$ we 
recover type IIB String Theory  from M-theory. Another example could be  
${\cal M}=R^{9}\times S^1/\z_2$ from which we will obtain type I string 
theory.

%
%
In order to study the relation of M-theory and the closed string
theories we will analyze the particular compactifications of M-theory on  $S^1$
corresponding to type IIA strings, and $S^1\times S^1$ that recovers, 
after using $T$-duality, the type IIB case. Open strings and the Heterotic 
case will be studied later.

In the IIA case the theory is defined by the radius of the  $S^1$, that we
will call $R$. The free energy  reduces to 
\be
\Lambda_{\r^{10}\times S^1}(\beta)  
\sim -\frac{V^T_{11} }{l_{11}^{10}}\int\frac{ds}{s^{6}}\,
\lc \theta_{3} \lp0,\frac{i\beta^2}{2\pi l_{11}^2 s}\rp - 
\theta_{4} \lp 0,\frac{i\beta^2}{2\pi l_{11}^2 s} \rp \rc 
 \theta_3\lp 0,\frac{i l_{11}^2  s}{2\pi R^2}\rp
\label{r10s1}
\ee
where  $l_{11}$ is the Planck length
in eleven dimensions. We want to compare this  magnitude 
to the Helmholtz free energy of the massless spectrum of type IIA String Theory  
and finally check that
\be
\Lambda_{\r^{10}\times S^1}(\beta)=\Lambda_{IIA}(\beta_{IIA}).
\ee
In the type IIA String Theory  the $0$-brane partons have masses given by
$\tilde{M}_n=n/\sqrt{\alpha'}g_s$, the tilde indicates we are measuring it in
ten-dimensional units. We can include it in the free energy simply by
putting this mass into string propagator. The free energy of the IIA string 
theory with D-partons  is given by
\be 
\Lambda_{IIA}(\beta_{IIA})\sim -\frac{V_{10}}{l_s^{10}} \sum_{N \in \z}\int\frac{ds}{s^{6}}\,
\lc \theta_{3} \lp0,\frac{i\beta_{IIA}^2}{2\pi \alpha' s }\rp - \theta_{4} \lp
0,\frac{i\beta_{IIA}^2}{2\pi \alpha' s} \rp \rc  e^{-\alpha' \tilde{M}_n^2 s}
{\cal Z}(s)                   
\label{IIA}.
\ee
where ${\cal Z}(s)$ is the internal partition function of the theory.
 
To establish the relationship between the theories we must measure the ten
dimensional temperature $\beta_{IIA}$ in M-theory  units. Using standard
Kaluza-Klein reduction it is easy to write the metrics $G_{D=10}$ in terms of
$G_{D=11}$. In the case we are describing we know that the 
metrics are also related with the string coupling constant as $G_{10}=g_s^{2/3}G_{11}$.
Then by using the relation between the radius $R$ and the  string coupling
$g_s$, the reduction of metrics leads to
the following relation between lengths\footnote{It
also implies that $\frac{V_{10}}{l_s^{10}}=\lp \frac{R}{l_{11}} \rp^5
\frac{V^T_{11}}{l_{11}^{10}}$ } 
\be
\frac{L_{10}}{l_s}=\lp \frac{R}{l_{11}}\rp
^{1/2}\frac{L_{11}}{l_{11}}
\ee  
Using this, and
thinking about the inverse temperature as a length,  we can  
express the function in (\ref{IIA}) in terms 
of $\beta$ and $R$, where $\beta$ is the M-theory temperature.
On the other hand we have to keep in mind tht  the string scale  is
expressed as $\sqrt{\alpha'}=l_s=g_s^{-1/3}l_{11}$ in terms of eleven-dimensional 
parameters \cite{SEI,SEN}.
and finally we have
\be
\Lambda_{IIA}(\beta)\sim - \frac{V_{10}}{\alpha'^{5}}\sum_{n\in \z}\int\frac{ds}{s^{6}}\,
\lc \theta_{3} \lp0,\frac{i\beta^2}{2\pi  s  l_{11}^2} \frac{R}{l_{11}}\rp - \theta_{4} \lp
0,\frac{i\beta^2 }{2\pi s l_{11}^2} \frac{R}{l_{11}} \rp \rc
e^{-\frac{l_{11}^3 M_n^2 s}{R}} {\cal Z}(s).	
\label{M-IIA}
\ee
where we have let the multiplicative factors in a ten-dimensional fashion.

In order to relate this expression to M-theory free energy we have to make 
the dimensional reduction of the latter from eleven to ten dimensions. Before 
doing so, it is necessary to make an appropriate rescaling of  the proper time 
in the previous expression
\be
s=\frac{l_{11}^2}{l_s^2} t. 
\label{rescaling}
\ee
In this way, and using the relations above, we find that (\ref{M-IIA}) reads
\be
\Lambda_{IIA}(\beta)\sim - \frac{V_{11}^T}{l_{11}^{10}}\sum_{n\in\z}
\int\frac{dt}{t^{6}}\,\lc \theta_{3} \lp 0,\frac{i\beta^2}{2\pi t l_{11}^2} \rp 
-\theta_{4} \lp 0,\frac{i\beta^2 }{2\pi t l_{11}^2} \rp \rc
e^{-l_{11}^2 M_n^2 t} {\cal Z}(\frac{l_{11}^2}{l_s^2} t).
\label{final}
\ee
Now the D-particle masses are measured in eleven-dimensional units, so we
have $M_n=n/R$.The connection to the free energy of M-theory on the circle is simply
obtained by decoupling massive excitations of the string. It is possible to
do it without integrating out the D-branes, by taking $g_s$ to be finite in
such a way $l_s^{-1}>>(l_s g_s)^{-1}$. In this limit ${\cal Z}(t)$ tends to
the unit and we exactly recover the free energy in (\ref{r10s1}). As expected, 
we have recovered the mechanism that allows the
relation between strong coupling dynamics of String Theory  at low energies and 
eleven-dimensional supergravity \cite{various}. The whole KK spectrum in M-theory
corresponds to bound states at threshold of D-particles of IIA theory.

The  perturbative string physics is reached by taking
$R\rightarrow 0$ in (\ref{r10s1}).  In this region the contribution corresponding  
to the KK tower of momentum states, tends to one 
(one could see the same in the D-particle picture by simply taking a
vanishing coupling constant limit)) and we obtain the free energy of the 
ten-dimensional IIA supergravity.

Let us see how the standard Hagedorn condition of  String Theory  is
expressed in terms of eleven-dimensional parameters. The usual method that
allows the computation of the Hagedorn temperatures gives, for the free
energy in (\ref{final}),
\be
\beta_H R^{1/2} \propto l_{11}^{3/2}.
\ee
We finally see that, as guessed, the critical point of the string gas is
given in terms of $\lambda=\beta R^{1/2}$.

Some words about type IIB String Theory  are in need. We know that this
theory is related to IIA one by $T$-duality. It seems then that by
analyzing M-theory on ${\cal M}=\r^9 \times S^1 \times S^1$ and using the
duality on one of the circles we will recover IIB strings from M-theory
physics \cite{various,horava,POWI,hull}. The  thermal properties of this
theory are not affected by the $T$-duality because we do not change the
`quantum' circle of M-theory.  

Another interesting related result that comes from the study of M-theory on the
torus is related to the FLIP in the Matrix Model of strings \cite{DVV,BSE,motl}.
From a purely M-theoretic point of view there is nothing special in taking $R$ 
or $R_9$ as the fundamental parameter which finally fixes the string scales. 
The system itself will choose the states of membranes wrapped around the
smallest circle as the 'perturbative' ones. We will see that it is related
to $S$-duality transformation.
This fact allows us to say that after flipping, the thermal properties of 
matrix String Theory   will depend on $\beta R_9^{1/2}$, as we guessed in the 
introduction.

\subsection{M-theory on orbifolds}

We can now review how M-theory contains the low energy regime of string
theories with one space-time supersymmetry \cite{horava}, that is, Heterotic
and open string theories.
Type I and type I' string theories can be obtained as an orientifold of type 
IIB and type IIA closed string, both are related by $T$-duality
\cite{orbi1,orbi2}. 
This suggests that there should be an  M-theory background that describes
type I theories. In fact it is possible to recover open strings by taking  
${\cal M}=R^{9}\times S^1 \times S^1$, and orbifolding  one of the
circles \cite{horava}. In this way  type I String Theory  is obtained from 
eleven-dimensional physics by considering M-theory on ${\cal M}=R^{9}\times
S^1/\z_2 \times S^1$. The whole map of open string theories is 
\ber
IIA \lc ({\cal M}=R^{9}\times S^1) \times S^1 \rc \stackrel{T}{\longleftrightarrow}
IIB \lc ({\cal M}=R^{9}\times \tilde{S}^1) \times S^1\rc
\nonumber
\\
\stackrel{\z_2}{\longleftrightarrow}
IB \lc ({\cal M}=R^{9}\times \tilde{S}^1/\z_2) \times S^1
\rc\stackrel{T}{\longleftrightarrow}
IA \lc ({\cal M}=R^{9}\times S^1/\z_2) \times S^1\rc
\label{map1}
\eer
where $T$ stands for $T$-duality. In the previous diagram we have recovered
the connected string theories without touching the eleven-dimensional $S^1$.
We know that the four theories above present the same Hagedorn temperature
\cite{ALV,OSO1}.
In our philosophy this is due to fact that the scale of the ten-dimensional 
theory is governed by the common circle.

By using the symmetry that interchanges the orbifolded $S^1$ in M-theory it is
possible to recover the $E_8 \times E_8$ Heterotic string as the M-theory on
$R^{9}\times S^1 \times S^1/\z_2 $. $T$-duality allows the relation of this
theories to the $SO(32)$ Heterotic string. We can write another relation
\be
E_8\times E_8 \lc ({\cal M}=R^{9}\times S^1) \times S^1/\z_2 \rc
\stackrel{T}{\longleftrightarrow}
SO(32) \lc ({\cal M}=R^{9}\times S^1) \times \tilde{S}^1/\z_2 \rc.
\label{map2}
\ee
In these theories the scale is defined by $S^1/\z_2$, and they have a
different critical temperature as those in (\ref{map1}). On the other hand,
it is also known that both maps, (\ref{map1}) and (\ref{map2}), are related
by $S$-duality \cite{various,hull,dab,POWI,horava} (that is the flip)
\be
SO(32) \lc ({\cal M}=R^{9}\times S^1) \times \tilde{S}^1/\z_2
\rc \stackrel{S}{\longleftrightarrow}
IB \lc ({\cal M}=R^{9}\times \tilde{S}^1/\z_2) \times S^1 \rc.    
\label{het/I}
\ee
These two theories include low energy excitations of string theories that have 
different Hagedorn temperatures. From our approach
it is due to the type of eleven-dimensional compact dimension they are
compactified in. We will argue later that this property comes from the
shrinking of one of the dimensions of the orbifolded torus, and consequently
from which are the membrane degrees of freedom we freeze.

From an M-theoretic point of view the change of the critical temperature by
the $S$-duality in (\ref{het/I}) occurs in two limits of the same
compactification background. By keeping both radii to be finite it is
possible to see how the Hagedorn critical point varies along a continuous
path from one theory to the other. More on this subject will be said in the
final discussion.

\section{Hagedorn transition from M(atrix)-Theory  }

Before analyzing the Hagedorn problem in the Matrix String Model let us briefly 
review how the Matrix Model of M-theory  recovers string theories from a 
non-perturbative perspective, and how it behaves on the corresponding
backgrounds at finite temperature. 

\subsection{Thermodynamics of Matrix Theory   on string backgrounds}

For simplicity we can start with the model that describes the type IIA String Theory  from
Matrix Membranes \cite{motl,DVV,BSE}. It consists of a system of  $N$ 
D-particles on $\r^9 \times S^1(R)\times S^1(R_9)$. Strings are related to
those configurations that correspond to two-dimensional membranes wrapped on
the torus. The radius $R$ is taken to be the light-cone one while  $R_9$ is 
the radius of the transverse spatial circle. On this background $N$
D0-branes, composing the wrapped membrane, are described by the 
dimensional reduction down to $1+1$ dimension of SYM initially in $D=9+1$ 
with  ${\cal N}=1$ and gauge group $U(N)$. 
The spatial dimension of the SYM theory is compactified on the $T$-dual circle 
$S^1(\tilde{R}_9)$ \cite{taylor}.

At this moment we have two parameters that define,  at least classically,
the theory. These parameters are the compactification radii of the torus
${\bf T}^2$. The relevant physics here comes from the possible limits we can 
take on the moduli space. 

We can relate this system to ten-dimensional physics in two different ways. 
Following the standard knowledge of $T$-duality, the action of the theory we
have is simply the D1-brane action of the type IIB String Theory  where the 
D-strings are wrapped around the compact dimension. The ten-dimensional 
interpretation is simply recovered by taking  the parameter $R$ to be small. 

On the other hand  IIA String Theory  is obtained from this framework by making 
the mentioned 'FLIP', that is using the radius $R_9$ as the responsible for the
connection to ten-dimensional physics. Geometrically it can been seen as the
shrinking of one of the dimensions of the membrane we started with. In this
way, for scales larger than $l_{11}$ the membranes look like ten
dimensional strings. The quantity $N$ stands for the KK momentum of the strings.  

Some thermal properties of the Matrix Model on the torus were studied in
\cite{us}. In that work 
\footnote{In order to compare the expressions here to the results in \cite{us} one must be careful
with  the normalizations. Here, we are working with those of \cite{hethorava}}
 it was obtained that the loop contributions to the
partition function on the torus take the following form 
\be
{\cal Z}_n \simeq (2\pi)^{-2n}\lp \frac{\beta R}{l_{11}^2}\rp^{3n}
\ee
where $R$ is the light-cone direction radius
\footnote{ 
We should multiply ${\cal Z}_n$ by $exp({-\frac{\beta N}{2 R}})$ that comes
from the light-cone degrees of freedom \cite{us}. This  exponential term does 
not need to be flipped because it comes from a purely light-cone frame energy and in the
case of matrix strings it is recovered by the string's momentum  modes
around the compact dimension $X\sim X+2\pi R$. When computing the string
partition function, in the context of Matrix String Model it will be
mandatory to include the light-cone modes that  will
correspond with the exponential term in the above expression.}.

The Heterotic string has also been obtained by a compactification of the
Matrix Model \cite{BM,hethorava,dan,lowe,silv,rey1,rey2}. The background manifold in
this case is $R^9\times S^1(R_9)/\z_2 \times  S^1(R)$. The methodology is
the same as in the type II case so $E_8\times E_8$ is reached by permuting
the r\^{o}le of the radii, and then the relationship to perturbative ten
dimensional physics is obtained by taking the  orbifold radius to be infinitely
small. Type I theory is obtained by shrinking the circular dimension,
$S^1(R)$.

Let us be more explicit. The Matrix Model for Heterotic string is
constructed  starting with the so-called Heterotic Matrix Theory  
\cite{dan,silv,rey1}, that corresponds to the Matrix Model on a transverse
orbifold $S^1/\z_2$. As usual the longitudinal direction is compactified in 
the eleven-dimensional circle. 
The D-particle system in this background may be described in two ways. We
can use a SYM quantum mechanics with the appropriate orbifold projection, and
the  addition of extra fermions coming from the orbifold planes, or by
$T-$dualizing the orbifold direction we arrive at the dimensional
reduction of $D=9+1$ SYM theory to $1+1$ dimensions, where the spatial one
is compactified on a circle of radius $\Sigma$ \cite{hethorava,dan,silv,rey1}.
 The residual gauge group is reduced to $O(N)$, and the 
parameters of the theory are related to fundamental ones by
\be
\tilde{g}^2=\frac{R^2}{R_9 l_{11}^3} \hspace*{20mm}  \Sigma=\frac{(2\pi)^2 l_{11}^3}{RR_9}
\ee
where $g$ is the coupling constant of  the SYM theory. The thermal properties 
of this model are qualitatively the same than those of theory on the torus. The 
difference will only arise from the internal degrees of freedom, then we can state 
\be
{\cal Z}_n^{S^1/\z_2} \simeq  \lp \frac{\tilde{g}^2 \beta^3}{\Sigma}\rp^n = 
(2\pi)^{-2n}\lp \frac{\beta R}{l_{11}^2}\rp^{3n}.
\ee
The final step in order to connect to Matrix String models we have to make
the FLIP. It will be simply done by interchanging the radius $R$ by $R_9$ in
the previous results. We are now able to attempt the problem of the critical
behavior of the string gas. This is the scope of next sections.

\subsection{The Hagedorn transition}

The previous analysis  has opened to us a possibility of studying some
features of Hagedorn transitions in Matrix theory. 

It has been recently proposed that Hagedorn temperatures represent a
deconfining point in the D-particle picture of perturbative strings. When the
temperature reaches the string scale the D-partons glued together in the
membrane configuration separate and a gas of individual D-particles comes
out. We know that the Matrix String Models reproduce perturbative string theories 
when the radius $R_9$, which, by flipping, is related to
the string coupling constant, is taken to be zero. It is precisely in this regime 
where the Hagedorn transition does arise, so the scale $R_9$ has to be considered
as the order parameter of the phase transition. This is the picture 
qualitatively sketched in  \cite{Sathi}.

In the previous sections we have argued that the critical properties of the 
string gases can be expressed in terms of the eleven-dimensional parameters, the
inverse temperature $\beta$ and the eleven-dimensional radius $R_9$ in  the
form of the 'thermal' parameter $\lambda=\beta R_9^{1/2}$. Let us see
how  this behavior comes in the Matrix Model of strings. We know that the
loop expansion at finite temperature is given by powers of $\beta
R_9/l_{11}^2$. Relating this magnitude to ten-dimensional variables, the
temperature and the scale $l_s$, and using the Hagedorn temperature, $
\beta_{IIA}^{H}$ as an input,  we arrive to the following Hagedorn condition 
for the eleven-dimensional temperature
\be
\frac{\beta_H R_9}{l_{11}^2}\propto g_s^{1/3} \Rightarrow \beta_H R^{1/2}
\propto l_{11}^{3/2}, 
\ee
as we obtained in section 2.

In this way it would be possible to state that
a unique value of this parameter, corresponding to the critical string
point $\lambda_H$, holds for various values of Matrix Theory temperature.  	

\section{Discussion}

In the previous section we have tried to extract some information from
M and Matrix theories about the thermal behavior of string effective
models. Finally we can conclude that, seen from eleven dimensions, the
ten-dimensional critical thermodynamics behaves as a function of 
$\lambda=\beta
R_9^{1/2}$, where $R_9$ is the radius of the eleventh dimension and it 
has to be taken as the order parameter of the Hagedorn phase transition. 

The first doubt we can elucidate is whether the Hagedorn temperature is
unique in the Matrix Model or if it depends on the compactification background. 
In perturbative String Theory  this uniqueness for each model, is supported
by the fact that the singularity is originated from the worldsheet degrees
of freedom. In the Matrix Model, if we suppose that the Hagedorn temperature is
unique for all the string models, we arrive at the conclusion that the
difference should come from compactification scales. Using the expression
(\ref{relation}) we see that the ratio $(R_9^{Het}/R_9^{IIA})^{1/2}$ should be
different than $1$. Remembering that
the string theories are obtained by taking the small $R_9$ limits, we can
finally conclude that the difference between closed type II and Heterotic
Hagedorn temperatures cannot come from the size of the 'quantum' radii.
This shows that it is expectable that in Matrix theory the Hagedorn
temperature depends on the compactification background. It will come from
the difference between reducing the internal degrees of freedom of the membrane,
shrinking the circle $S^1$, as in type II and I strings, or the orbifold 
$S^1/\z_2$ which corresponds to the Heterotic case.

Another important consequence of the thermal behavior we have obtained is 
that while the critical point in String Theory   is given in terms of a single 
value of $\lambda$, from a M-theoretical point of view it correspond to a 
family of $\beta$ and $R_9$ parameters. We can see that it is possible to
relate this property to $S$-duality. 

We will begin by focusing on the IIB case. This theory is obtained from
Matrix Theory   by $T$-dualizing  the compact dimension of radius $R$ of the
IIA matrix strings. We know 
that the Hagedorn condition  for type IIB closed strings, 
$\beta_{IIB}^{H}=\pi\sqrt{8\alpha'}$, leads to the following one for eleven 
dimensional temperature
\be
\beta^{H}=\sqrt{8} \pi \lp \frac{l_{11}^3}{R_9} \rp^{1/2}.   
\label{hag1}
\ee
On the other hand we can introduce the thermal properties of
non-perturbative effects in the same fashion. We include the degrees of freedom of D-strings in
type IIB theory  by taking the $R\rightarrow 0$ limit on
the Matrix Theory   on $S^1(R_9)\times S^1(R)$. In this case $R$ defines the
string scales so the thermal parameter is $\lambda'=\beta_D R^{1/2}$. The critical 
temperature, $\beta_D^{H}$ of this gas will be the same of (\ref{hag1}) but
with $R$ instead of $R_9$. 
The relation between the critical temperatures is given by
\be
\beta^{H} \lp \frac{R_9}{R}\rp^{1/2}=\beta^{H} \lp \frac{g_s}{g_D} \rp
^{1/3}=\beta_D^{H} 
\ee
We know that in the picture we are working $R_9$ is taken to be smaller than
$R$ in such a way the non-perturbative effects present a critical
temperature bigger than fundamental strings. This property is an explanation
of a known behavior presented in \cite{barbovaz}. We have related it to
$S$-duality by expressing the radii $R_9$ and $R$ in terms of two 'string' 
coupling constants.  
We easily see that when the fundamental string coupling is small, the gas of
non perturbative strings arrive at its critical temperature after the 
Hagedorn transition of the perturbative ones. The contrary happens for strong 
coupling dynamics where  the D-string gas has to be taken as the perturbative 
one. 

In the case of the Heterotic/type I duality the picture is practically the same.
The main difference comes from the Hagedorn condition of string models. We
know that the Heterotic gas presents its Hagedorn temperature at
$\beta_{Het}^{H}=(2+\sqrt{2})\pi\sqrt{\alpha'}$. The $S$-dual gas of this 
theory is composed of open strings whose critical
point is the same as in Type II theories.
The relationship in terms of coupling constant, analogous to that  given
for the IIB string model, is for the case of Heterotic/Type I duality
\be
\beta_{Het}^{H} \lp \frac{g_{Het}}{g_{open}} \rp ^{1/3}= \frac{\lp \sqrt{2}+1
\rp }{2} \beta_{open}^{H}.
\label{s-dualhet}
\ee 
Remembering that by $S$-duality the coupling constant are related as
$g_{open}=1/g_{Het}$, and equivalently $g_s=1/g_D$, we can make a general
statement that relates critical temperatures of $S$-dual gases
\be
\beta^{H} g^{2/3} = a \beta_{Dual}^{H}
\ee
where $a$ is different for the various cases of $S$-dual theories. This
property seems to be an evidence for the guess in \cite{BM}, in the sense
that the theories with the same number of supersymmetries are continuously
related while it seems that there is not a continuous connection to theories
with a different amount of supersymmetry.

\section{Acknowledgments}
We thank M.A.R. Osorio for useful conversations. The work of M.Laucelli
Meana is supported by M.E.C. under FP97 grant.

\end{document}